\documentclass[a4paper,fleqn,usenatbib]{mnras}

\usepackage{newtxtext,newtxmath}

\usepackage[T1]{fontenc}
\usepackage{ae,aecompl}
\usepackage{epstopdf}

\usepackage{graphicx}	
\usepackage{amsmath}	
\usepackage{amssymb}	
\usepackage{bm}

\title[Non-isothermal vortex evolution]{On the vortex evolution in non-isothermal protoplanetary discs}

\author[D. Tarczay-Neh\'ez, Zs. Reg\'aly \& E. Vorobyov]{D. Tarczay-Neh\'ez$^{1,2}$, Zs. Reg\'aly$^{1}$ and E. Vorobyov$^{3,4,5}$
\\
$^{1}$Konkoly Observatory, Research Centre for Astronomy and Earth Science, Konkoly-Thege Mikl\'os 15-17, 1121, Budapest, Hungary\\
$^{2}$MTA CSFK Lend\"ulet Near-Field Cosmology Research Group\\
$^{3}$Department of Astrophysics, University of Vienna, 1180, Vienna, Austria\\
$^{4}$Research Institute of Physics, Southern Federal University, Stachki Ave. 194, 344090, Rostov-on-Don, Russia\\
$^{5}$Ural Federal University, 51 Lenin Str., 620051 Ekaterinburg, Russia}

\date{ Accepted 2020 February 5. Received 2020 February 5; in original form 2019 August 2}

\pubyear{2020}

\begin{document}
\label{firstpage}
\pagerange{\pageref{firstpage}--\pageref{lastpage}}
\maketitle

\begin{abstract}
It is believed that large-scale horseshoe-like brightness asymmetries found in dozens of transitional protoplanetary discs are caused by anticyclonic vortices. These vortices can play a key role in planet formation, as mm-sized dust -- the building blocks of planets -- can be accumulated inside them. Anticyclonic vortices are formed by the Rossby wave instability, which can be excited at the gap edges opened by a giant planet or at sharp viscosity transitions of accretionally inactive regions. It is known that vortices are prone to stretching and subsequent dissolution due to disc self-gravity for canonical disc masses in the isothermal approximation. To improve the hydrodynamic model of protoplanetary discs, we include the disc thermodynamics in our model. In this paper, we present our results on the evolution of the vortices formed at the outer edge of an accretionally inactive region (dead zone) assuming an ideal equation of state and taking $PdV$ work, disc cooling in the $\beta$-approximation, and disc self-gravity into account. Thermodynamics affects the offset and the mode number (referring to the number of small vortices at the early phase) of the RWI excitation, as well as the strength, shape, and lifetime of the large-scale vortex formed through merging of the initial small vortices. We found that the inclusion of gas thermodynamics results in stronger, however decreased lifetime vortices. Our results suggest that a hypothetical vortex-aided planet formation scenario favours effectively cooling discs.
\end{abstract}




\section{Introduction}
\label{sec:intro}

Accretion discs are the places of origin of planets and planetary systems around an embedded young stellar object. Discs are evolving from the primordial gas- and dust-rich phase to the gas-poor and (second generation) dust-rich debris disc phase, in-between the transitional phase where gas is significantly depleted. In the recent several years, inner cavities and large-scale horseshoe-like lopsided morphologies were found in the millimetre-wavelength images in dozens of transitional discs (see e.g. \citealp{Andrewsetal2009,Andrewsetal2011}; \citealp{Brownetal2009}; 
\citealp{Hughesetal2009}; \citealp{Isellaetal2010}; \citealp{Mathewsetal2012}; \citealp{Tangetal2012}; \citealp{Fukagawaetal2013}; \citealp{Casassusetal2013,Casassusetal2015}; \citealp{vanderMareletal2013}; \citealp{Perezetal2014}; \citealp{Hashimotoetal2014}; \citealp{Wrightetal2015}; \citealp{Momoseetal2015}; \citealp{Marinoetal2015}). These brightness asymmetries are thought to be caused by dust accumulation in large-scale anticyclonic vortices. In the centre of an anticyclonic flow, pressure maximum takes place in which dusty material can be trapped (see e.g. \citealp{AdamsandWatkins1995}; \citealp{BargeandSommeria1995}; \citealp{Tangaetal1996}; \citealp{KlahrandHenning1997}; \citealp{Braccoetal1999}; \citealp{GodonandLivio2000}), therefore these formations can play a major role in planet formation (see e.g. \citealp{KlahrandBodenheimer2006}; \citealp{HengandKenyon2010}; \citealp{OwenandKollmeier2017}). 

There are a few other phenomena which can explain such morphologies, e.g., disc eccentricity excited by a massive companion star \citep{Ragusaetal2017}, self-shadowing caused by a tilted inner disc due to an inclined giant planet \citep{DemidovaandGrinin2014} or at the outer edges of the accretion disc due to mass flow from the natal cloud \citep{Baeetal2015}. However, in this paper, we only focus on the Rossby Wave Instability excited by vortices.

Two-dimensional hydrodynamical numerical studies of protoplanetary discs suggest that these large-scale vortex formations in Keplerian accretion discs can be excited by the non-axisymmetric hydrodynamic instability, the so-called Rossby Wave Instability \citep{Rossby1939}, hereafter referred as RWI (e.g. \citealp{Lovelaceetal1999}; \citealp{KlahrandBodenheimer2003}; \citealp{LyraandKlahr2011}; \citealp{Raettingetal2013}; \citealp{Lyra2014}). The RWI is excited at the vortensity extremum (minimum), which can evolve at a steep pressure gradient in protoplanetary discs. Such features can occur at the edges of a gap opened by an embedded massive planet \citep{Lietal2005}, or at the edges of the accretionally inactive zone of discs (\citealp{VarniereandTagger2006}; \citealp{Lyraetal2009b}; \citealp{Meheutetal2010,Meheutetal2012a,Meheutetal2012b,Meheutetal2012c,MeheutetandLai2013}; \citealp{Crespeetal2011}; \citealp{Regalyetal2012}; \citealp{Richardetal2013} and \citealp{Flocketal2015}). 

The lifetime of an anticyclonic vortex in a protoplanetary disc is a crucial point in planet formation. Recent studies confirm that disc viscosity can reduce the strength and the lifetime of the vortices formed at gap edges opened by a massive planet (\citealp{devalBorroetal2006};
\citealp{Ataieeetal2013}; \citealp{Fuetal2014b}; \citealp{Mirandaetal2016}). Dust can accumulate in the vicinity of the centre of a long-lived vortex. Nevertheless, if the dust-to-gas mass ratio reaches unity, the impact of dust on gas (dust feedback) also plays a major role in destroying the vortex on a local scale (\citealp{Johansenetal2004}; \citealp{InabaandBarge2006}; \citealp{Lyraetal2009a}; \citealp{Fuetal2014a}; \citealp{Crnkovicetal2015}; \citealp{Survilleetal2016}). \cite{Raettingetal2015} found that this feedback causes vortices to disappear and re-build again with time. Recently, \cite{Mirandaetal2017} revealed that the process of azimuthal dust trapping slows down due to dust feedback in the circumstance of large-scale vortices formed at viscosity transitions. Another effect in high-mass discs which can damp or delay the formation of large-scale vortices formed at the edge of a planet-opened gap is self-gravity (\citealp{LinandPapaloizou2011} and \citealp{Lin2012}). 

Vortices formed at artificial pressure bumps are found to be weakened by self-gravity of the disc \citep{ZhuandBaruteau2016}. \cite{Baeetal2015} found that large-scale vortices, developed in the outer regions of protostellar discs due to mass-loading from natal clouds, dissipate, as the Toomre $Q$-parameter \citep{Toomre1694} reaches unity. \cite{LovelaceandHohlfeld2013} and \cite{Yellinetal2016} investigated disc self-gravity and found that it is important for discs with Toomre parameter $Q~<~Q_{\mathrm{crit}}~=~1/h$, where $h$ denotes to the geometry (apsect ratio) of the disk. Recently, \cite{RegalyandVorobyov2017a} investigated the role of gas self-gravity in the long-time evolution of vortices at sharp viscosity transitions in protoplanetary discs. They found that at relatively low-mass discs ($M_{\mathrm{disk}}/M_{\star}~\geq~0.006$), self-gravity becomes an important effect as it stretches, azimuthally elongates, and weakens the RWI formed vortices and shorten their lifetimes.

Recently, \cite{PierensandLin2018} investigated the effect of thermodynamics on the long-term evolution of RWI-excited vortices at the inner boundary of the dead zone. They used both the $\beta$-prescription and black-body cooling to simulate the effect of thermodynamics. They found that in a non-self-gravitating case, increasing $\beta$ strengthens the vortex. They also investigated the case where self-gravity is included. They found that with low disc masses, the effect of $\beta$-cooling is the same as in non-self-gravitating models, and the results of $\beta\leq0.1$  are consistent with the locally isothermal simulations. At higher disc masses, \cite{PierensandLin2018} found that disc self-gravity becomes dominant over $\beta$-cooling driven thermodynamics. However, with an assumption of black body cooling, self-gravity can stabilise vortex. Here we note that they investigated models with $\beta~=~0.01$ to $1$.

In this paper, we examine the role of the thermodynamics on vortex formation, evolution and lifetime. In Section \ref{sec:hydro}, we give a short description of our 2D hydrodynamic model used for the simulations. In Section \ref{sec:res} we show our results about the effect of thermodynamics on vortex formation at the edge of an accretionally inactive region. In Sections \ref{sec:discussion} and \ref{sec:concl} we give a discussion and conclusion about our results.

\section{Hydrodynamic model}
\label{sec:hydro}

We investigate the long-term evolution of large scale vortices formed in protoplanetary disc by means of two-dimensional hydrodynamical simulations. For this investigation, we use an extension of the \textsc{gfargo} code\footnote{http://fargo.in2p3.fr/-GFARGO-}, which is a GPU supported version of  \textsc{fargo} \citep{Masset2000}. \textsc{gfargo} numerically solves the vertically integrated continuity and the Navier--Stokes equations on a 2D polar ($R,\phi$) grid. 

In this investigation, we compare the vortex evolution in a locally isothermal approximation to a non-isothermal one. For the latter, the energy conservation has to be solved, thus we implemented a numerical solver module based on the \textsc{zeus} hydro code \citep{StoneandNorman1992}. Our implementation takes into account cooling and heating term for the energy conservation equation.

\subsection{Hydrodynamic equations}
\label{sec:thermo}

The continuity, Navier-Stokes and energy conservation equations that govern the protoplanetary disc dynamics read as
\begin{equation}
    \frac{\partial \Sigma}{\partial t}+\nabla \cdot (\Sigma\bm{v})=0,
\label{eq:cont}
\end{equation}
\begin{equation}
    \frac{\partial \bm{v}}{\partial t}+(\bm{v}\cdot \nabla)\bm{v}=-\frac{1}{\Sigma} \left( \nabla P+ \nabla{\cdot T} \right) -\nabla \Phi_\mathrm{tot},
\label{eq:NS}
\end{equation}
\begin{equation}
    \frac{\partial{e}}{\partial t}+\nabla \cdot (e \bm{v})=-P\nabla{\cdot \bm{v}}+Q_\pm,
\label{eq:energy}
\end{equation}

\noindent where $\Sigma$, $\bm{v}$, $P$, and $e$ are the surface mass density, velocity, vertically integrated pressure and thermal energy density of the gas (per surface area), respectively, and $T$ is the viscous stress tensor. $\Phi_{\mathrm{tot}}$ is the total gravitational potential of the central star ($\Phi_{\star}$), the disc itself ($\Phi_{\mathrm{sg}}$) and the indirect potential ($\Phi_{\mathrm{ind}}$), which appears due to the fact that the origin of the grid is not the barycentre of the disc (for more details see e.g. \cite{MittalandChiang2015}, \cite{ZhuandBaruteau2016} and \cite{RegalyandVorobyov2017b}):

\begin{equation}
 \Phi_{\mathrm{tot}} = \Phi_{\star} + \Phi_{\mathrm{ind}} + \Phi_{\mathrm{sg}},
\end{equation}

\noindent where $\Phi_{\star}$,  $\Phi_{\mathrm{sg}}$ and $\Phi_{\mathrm{ind}}$ at a given  distance, $r$, can be given as

\begin{equation}
\Phi_{\star} = -G\frac{M_{\star}}{r},
\end{equation}
\begin{equation}
\Phi_{\mathrm{ind}} = r\cdot G \int{\frac{\mathrm{d}m(\bm{r'})}{r^3}\bm{r'}},
\end{equation}
\begin{equation}
\label{eq:phisg}
\Phi_{\mathrm{sg}} = -G \int_{r_{\mathrm{in}}}^{r_{\mathrm{out}}}{r'dr} \times \int_0^{2\pi}{\frac{\Sigma\mathrm{d}\Phi'}{\sqrt{r'^2 + r^2 - 2rr' \cos{ (\Phi'-\Phi )}}}}.
\end{equation}

Here $r_{\mathrm{in}}$ and $r_{\mathrm{out}}$ are the inner and outer boundaries of the disk, and $m$ is the mass contained in a given grid cell. Equation~(\ref{eq:phisg}) is solved by Fast Fourier Transform technique \cite[see details in Section~2.8 in][]{BinneyandTremaine1987}, which was successfully applied to investigate fragmentation in gravitationally unstable protoplanetary discs (see, e.g., \citealp{VorobyovandBasu2010}, \citeyear{VorobyovandBasu2015}) and vortex formation in self-gravitating discs (see, e.g., \citealp{RegalyandVorobyov2017a}). Note that, according to \cite{RegalyandVorobyov2017a}, no gravitational softening is applied to our simulations, which may influence the effect of self-gravity.

The angular momentum transport due to turbulent viscosity is modelled by the $\alpha$--prescription of \cite{ShakuraandSunyaev1973}. In order to excite the RWI, which leads to the formation of vortices, we introduced an accretionally inactive region, known as the \emph{dead zone} \citep{Gammie1996}, by reducing the kinematic viscosity of the gas by a factor $\delta_{\alpha}$
\begin{equation}
\label{eq:viscred}
\delta_{\alpha} = 1 - \frac{1}{2}\left ( 1 - \alpha_{\mathrm{mod}}  \right ) \left [1 - \tanh{\left ( \frac{r - r_{\mathrm{dze}}}{\Delta r_{\mathrm{dze}}} \right )} \right]
\end{equation}
at a certain distance (see, e.g., \citealp{Regalyetal2012}). In Equation~(\ref{eq:viscred}) $\alpha_{\mathrm{mod}}$ is the depth of the reduction, $r_{\mathrm{dze}}$ and $\Delta r_{\mathrm{dze}}$ are the location and the half-width of the viscosity reduction. With Equation~(\ref{eq:viscred}) the global viscosity of the gas is $\alpha \delta_{\alpha}$. In this way, we only model the outer edge of the dead zone. The $\alpha$-parameter is set to $10^{-2}$, and $\alpha_{\mathrm{mod}}~=~10^{-3}$ is used in all models.

In a locally isothermal approximation (hereafter referred as I), thermal heating and cooling processes occur on time-scales that are much faster than the local dynamical timescale and the disc heating sources do not vary appreciably on time-scales of interest. The former is usually fulfilled on radial distances greater than a few AU \citep[see Fig.~3 in][]{Vorobyov2014} and the latter is true if stellar luminosity and viscous heating vary weakly with time. To solve Equations~(\ref{eq:cont})-(\ref{eq:energy}), one needs the equation of state of the gas, which takes the form $P_{\mathrm{I}}~=~\Sigma c_{\mathrm{s, I}}^2$, where the locally isothermal sound speed reads ($c_{\mathrm{s,I}}$) as

\begin{equation}
c_{\mathrm{s, I}}~=~H\Omega,
\label{eq:csi}
\end{equation}

\noindent$H~=~hr$ is the local scale-height, $h$ describes the geometry of the disc (aspect ratio), $\Omega=\sqrt{GM_{\star}/r^{3}}$ is the angular velocity, and $G$ and $M_{\star}$ are the gravitational constant and the stellar mass, respectively (set to unity). 
 
In a locally isothermal approximation, the temperature of the gas, $T_{\mathrm{g}}~=~c_{\mathrm{s,I}}^2 \mu / \mathcal{R}$ (where $\mu$ is the mean molecular mass and $\mathcal{R}$ is the universal gas constant, both set to unity in the applied numerical code), depends only on the distance from the star. This approximation, therefore, neglects the heating and cooling caused by the interaction of disc and vortices. This might be unrealistic because vortices excite spiral shock waves and interact with the disk.

In non-isothermal models (hereafter referred as NI), a more realistic description of the thermal processes is assumed. In these models, Equations~(\ref{eq:cont})-(\ref{eq:energy}) are closed with the ideal equation of state
\begin{equation}
  \label{eq:state}  
  P=(\gamma-1)e,
\end{equation}
with the the adiabatic index $\gamma=1.4$, also known as the polytropic index. With this approximation, we take into account the heating and cooling effects due to the expansion and compression of the gas, i.e., due to the $P \mathrm{d}V$ work. 
We also take into account various possible cooling or heating mechanisms implicitly via $Q_\pm$, such as stellar and background irradiation, and dust cooling, which can take the disc back to the thermal equilibrium state. For that, we used the $\beta$-cooling/heating prescription to let the gas release/gain its internal energy. According to \cite{LesandLin2015},
\begin{equation}
\label{eq:qparam}
    Q_\pm = \frac{1}{\tau_c}\left ( e - e^0\frac{\Sigma}{\Sigma^0} \right ),
\end{equation}
\noindent where $\tau_c$ is the cooling time connected with the $\beta$-parameter as
\begin{equation}
\label{eq:tauc}
 \tau_c = \frac{\beta}{\Omega}.
\end{equation}

In Equation (\ref{eq:qparam}), $\Sigma^0$ and $e^0$ correspond to the initial density and energy state of the disk. The sound speed is given as $c_{\mathrm{s}}~=~\sqrt{\gamma P/\Sigma}$, therefore, in the non-isothermal approximation the sound speed can be expressed as
\begin{equation}
\label{eq:csni}
    c_{\mathrm{s, NI}}~=~\sqrt{\frac{\gamma \left( \gamma -1 \right ) e }{\Sigma}}
\end{equation}
using Equation~(\ref{eq:state}). 

To investigate the effect of thermodynamics, we run simulations with two different approaches. First, we assumed that aspect ratio of the non-isothermal and locally isothermal discs are the same (Type~I simulations). By assuming same initial pressure and density distributions in locally isothermal and non-isothermal Type~I models, the initial sound speed in the locally isothermal ($c_{\mathrm{s, I, TI}}^0$) and non-isothermal ($c_{\mathrm{s, NI, TI}}^0$) case are different
\begin{equation}
    \label{eq:css}
    c_{\mathrm{s, NI, TI}}^0~=~\sqrt{\gamma} c_{\mathrm{s, I, TI}}^0,
\end{equation}
where $\gamma~=~1$ in the locally isothermal case. In the locally isothermal case, $c_{\mathrm{s,I}}$ is constant in time, while in the non-isothermal case, $c_{\mathrm{s,NI}}$ depends on $e$ and $\Sigma$, thus evolving in time. Emphasise that while Equation\,(\ref{eq:csni}) holds at any time of the simulations, Equation\,(\ref{eq:css}) is fulfilled only at the beginning of the simulations. 

According to the \cite{ShakuraandSunyaev1973} prescription, $c_{\mathrm{s}}$ affects the kinematic viscosity of the gas, thus the evolution of gas via the Navier-Stokes equations (see Equation~(\ref{eq:NS})). Hence, in the non-isothermal case, the evolution of the pressure gradient is influenced by the evolution of the kinematic viscosity, which is connected with $\beta$ via the energy density ($e$). In Type~II simulations, we assumed that the initial temperature distribution of the non-isothermal and locally isothermal simulations are the same. As we used $\alpha$-prescription, this means that the initial aspect ratios of the non-isothermal ($h_{\mathrm{I}}^0$) and locally isothermal ($h_{\mathrm{NI, TII}}^0$) simulations are different
\begin{equation}
    \label{eq:h}
    h_{\mathrm{I}}^0 = h_{\mathrm{NI, TII}}^0 / \sqrt{\gamma}.
\end{equation}

We also compared the locally isothermal and non-isothermal models assuming three different disc masses in the self-gravitating and non self-gravitating limits.

\begin{table*}
\caption{Used parameters for simulations.}
\label{tab:example}
\begin{tabular}{lcccccccc}
\hline\hline
Simulation & self-gravity & $h^0$ & $c_{\mathrm{s}}^0$ &  $r_{\mathrm{dze}}$ & $\Delta r_{\mathrm{dze}}$ & $\Sigma$ & $M_{\mathrm{disk}}$& $\beta$\\
& & & & [AU] & [AU] &  [$M_{\sun}/\mathrm{AU^2}$] &[$M_{\star}$]\\
\hline\hline\noalign{\smallskip}
\multicolumn{8}{c}{\footnotesize{\emph{Locally isothermal simulations}}}\\
\noalign{\smallskip}Locally isothermal (I) & YES & $h_{\mathrm{I}}$ & $c_{\mathrm{s,I}}$ & 24 & 1.68  & $3.18099 \cdot 10^{-6}$ & 0.001 & --\\
   & NO &&& && $1.909589 \cdot 10^{-5}$ & 0.006 & \\
&&    &   &  & &  $3.18099 \cdot 10^{-5}$ & 0.01 \\

\hline\hline\noalign{\smallskip}
\multicolumn{8}{c}{\footnotesize{\emph{Non-isothermal simulations}}}\\
\noalign{\smallskip}
Non-isothermal (NI) & YES & $h_{\mathrm{NI}}^0~=~h_{\mathrm{I}}$ & $c_{\mathrm{s,NI}}^0~=~\sqrt{\gamma}c_{\mathrm{s,I}}$ & 24 &1.68 & $3.18099 \cdot 10^{-6}$ & 0.001 & 0.1\\
(Type~I)    & NO  && && & $1.909589 \cdot 10^{-5}$ & 0.006 & 0.3\\
    &   &  &&& &  $3.18099 \cdot 10^{-5}$ & 0.01 & 1\\
    &   &  && &&   &   &   3 \\
    &   &   &&&&   &   &   10\\

\hline
\noalign{\smallskip}Non-isothermal (NI) & YES& $h_{\mathrm{NI}}^0~=~h_{\mathrm{I}}/\sqrt{\gamma}$ & $c_{\mathrm{s,NI}}^0~=~c_{\mathrm{s,I}}$ & 24 & 1.49  & $3.18099 \cdot 10^{-6}$ & 0.001 & 0.1\\
(Type~II)    & NO &&&& & $1.909589 \cdot 10^{-5}$ & 0.006 & 0.3\\
    &   &   &  &&&$3.18099 \cdot 10^{-5}$ & 0.01 & 1\\
    &   &   &   &&&&   &   3 \\
    &   &   &   &  &&& &   10\\

\hline\hline
\end{tabular}
\end{table*}

\subsection{Initial and boundary conditions}
\label{sec:initial}

The so-called flat-disc approximation is used, in which case the disc aspect ratio is assumed to be $h~=~0.05$ both in locally isothermal and $\beta$-cooling Type~I simulations. However, in Type~II the geometry of the disc is different ($h_{\mathrm{NI,TII}}~=~h/\sqrt{\gamma}$, see Section \ref{sec:thermo}). We used damping boundary conditions for $\Sigma$, $v_{\mathrm{r}}$, $v_{\mathrm{\phi}}$ and $e$ at the inner and open boundary conditions (see details in \citealp{devalBorroetal2006}) for the outer boundary of the computational domain. The inner and outer boundaries of the disc are set as $r_{\mathrm{in}}~=~3$ and $r_{\mathrm{out}}~=~50$~AU. The numerical resolution of the simulation domain is defined by logarithmically distributed $N_{\mathrm{r}}~=~256$ and equidistant $N_{\mathrm{\phi}}~=~512$  grid cells into radial and azimuthal directions, respectively. 
In addition, to verify numerical convergency, we run simulations with different numerical resolutions: $256\,\times\,512$, $512\,\times\,1024$ and $1024\,\times\,2048$. All other parameters were unchanged. Comparing the results, we conclude that our simulations are in the numerically convergent regime with the applied numerical resolution.

The width of the viscosity transition region is set to $\Delta r_{\mathrm{dze}}~=~1.4$~H. According to \cite{MatsumuraandPudritz2005} the viscosity radius of reduction is set to $r_{\mathrm{dze}}~=~24$~AU. It means that the half-width of the viscosity transition region equals to $1.68$~AU locally isothermal and in Type~I non-isothermal, while $1.49$~AU in Type~II non-isothermal simulations. 

Initially, the surface density of the gas is written as a power-law function: $\Sigma~=~\Sigma_0r^{-a}$, where $\Sigma_0$ is surface density at $r~=~1$AU, and $a~=~1$ is the power-law index of the surface density profile. \cite{RegalyandVorobyov2017a} and \cite{ZhuandBaruteau2016} have shown that disc mass affects the life-time of vortices assuming locally isothermal approximation. Therefore we investigate non-self-gravitating and self-gravitating models assuming three disc masses ($M_{\mathrm{disk}}/M_{\star}\,=\,0.001$,\,$0.06$ and $0.01$), with three corresponding $\Sigma_0$ values of $3.18099\cdot10^{-6}$, $1.909859\cdot10^{-5}$ and $3.18099\cdot10^{-5} \mathrm{M_{\odot}/AU^2}$.

Five different $\beta$--cooling prescription cases are investigated: $\beta\,=\,0.1$,\,$0.3$,\,$1$,\,$3$ and $10$. Based on Equation (\ref{eq:qparam}) $\beta$ refers to the number of orbit at a given radius required to settle the gas energy to the initial state.

\begin{figure*}
  \includegraphics[width = \textwidth]{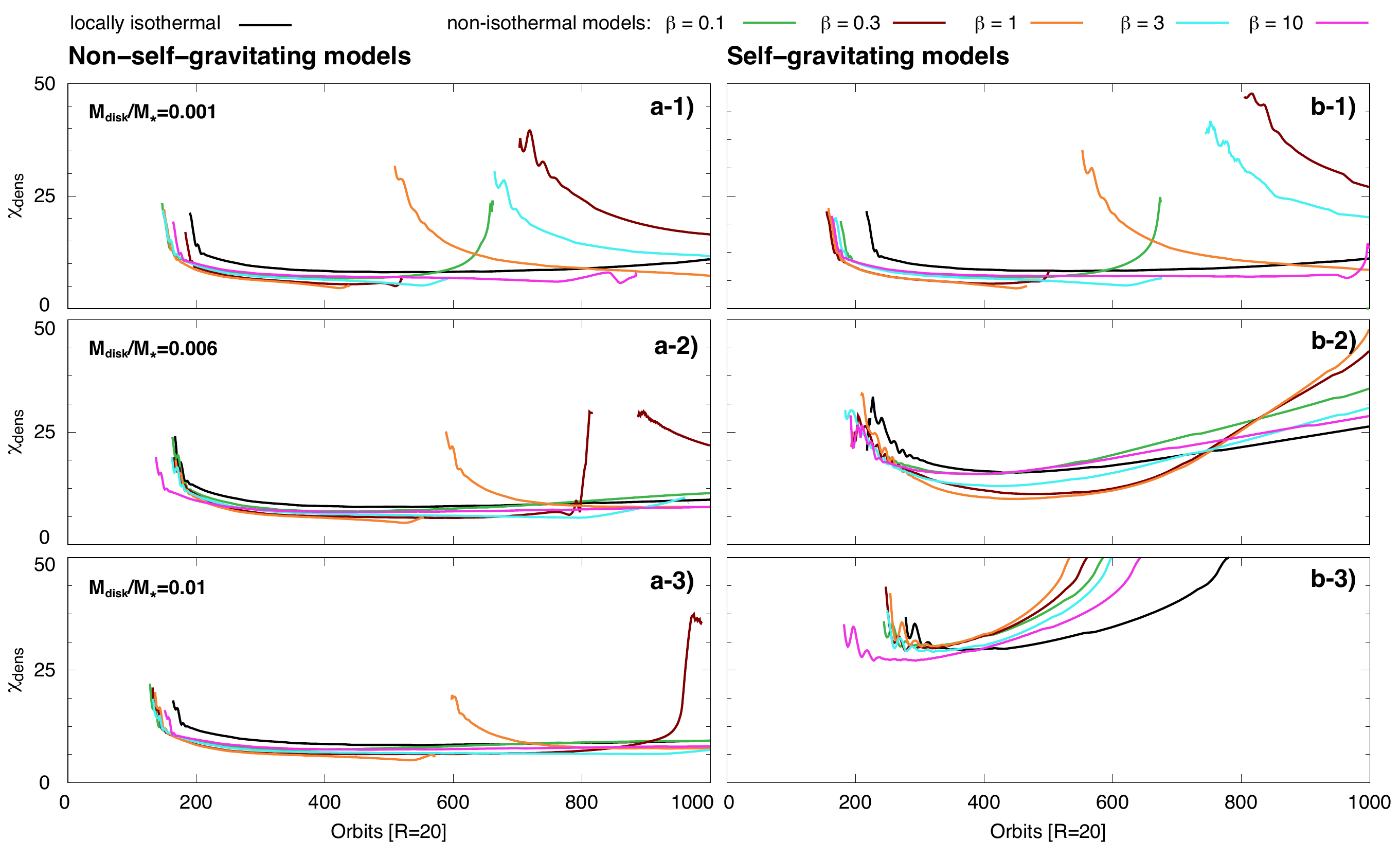}
    \caption{The evolution of vortex aspect ratio, $\chi_\mathrm{dens}$, as a function of time measured in the number of Keplerian orbital period, at the distance of the vortex eye in non-self-gravitating (left-hand-side panels) and self-gravitating Type~I models (right-hand-side panels) assuming three different disc masses ($M_{\mathrm{disk}}/M_{\star}~=~0.001,~0.006$ and $0.01$ from top to bottom). The black line refers to the locally isothermal case, while the coloured lines refer to the non-isothermal cases.}
    \label{fig:chi_contr_e-5}
\end{figure*}

\section{Results}
\label{sec:res}

\begin{figure}
    \centering
    \includegraphics[width = \columnwidth]{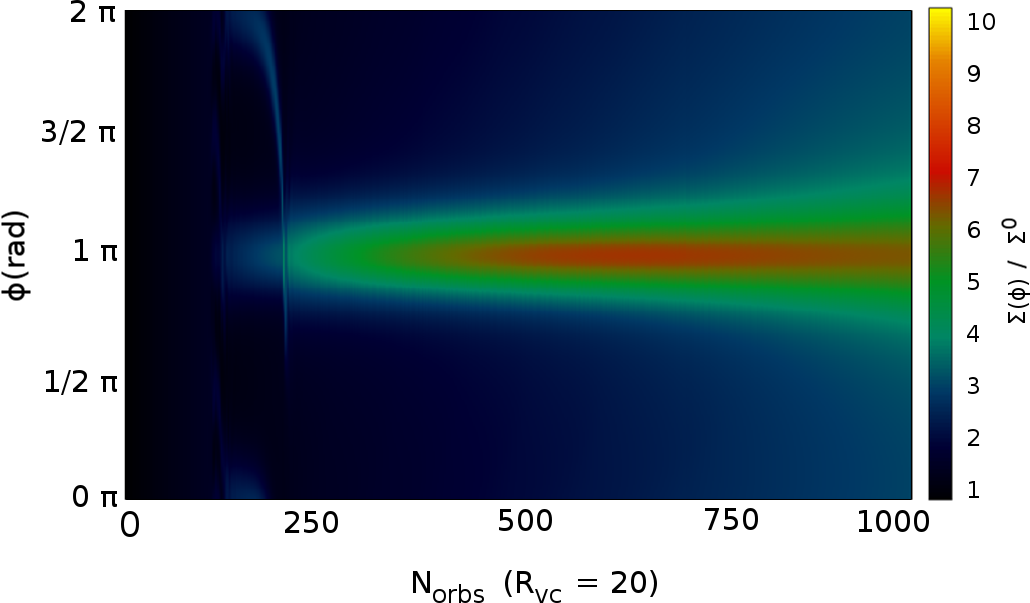}
    \caption{Vortex azimuthal density profile time evolution (in units of vortex orbit) without self-gravity in the case of low disc mass Type~II model with $\beta~=~0.01$.}
    \label{fig:lowbeta}
\end{figure}

\begin{figure*}
    \centering
    \includegraphics[width = \textwidth]{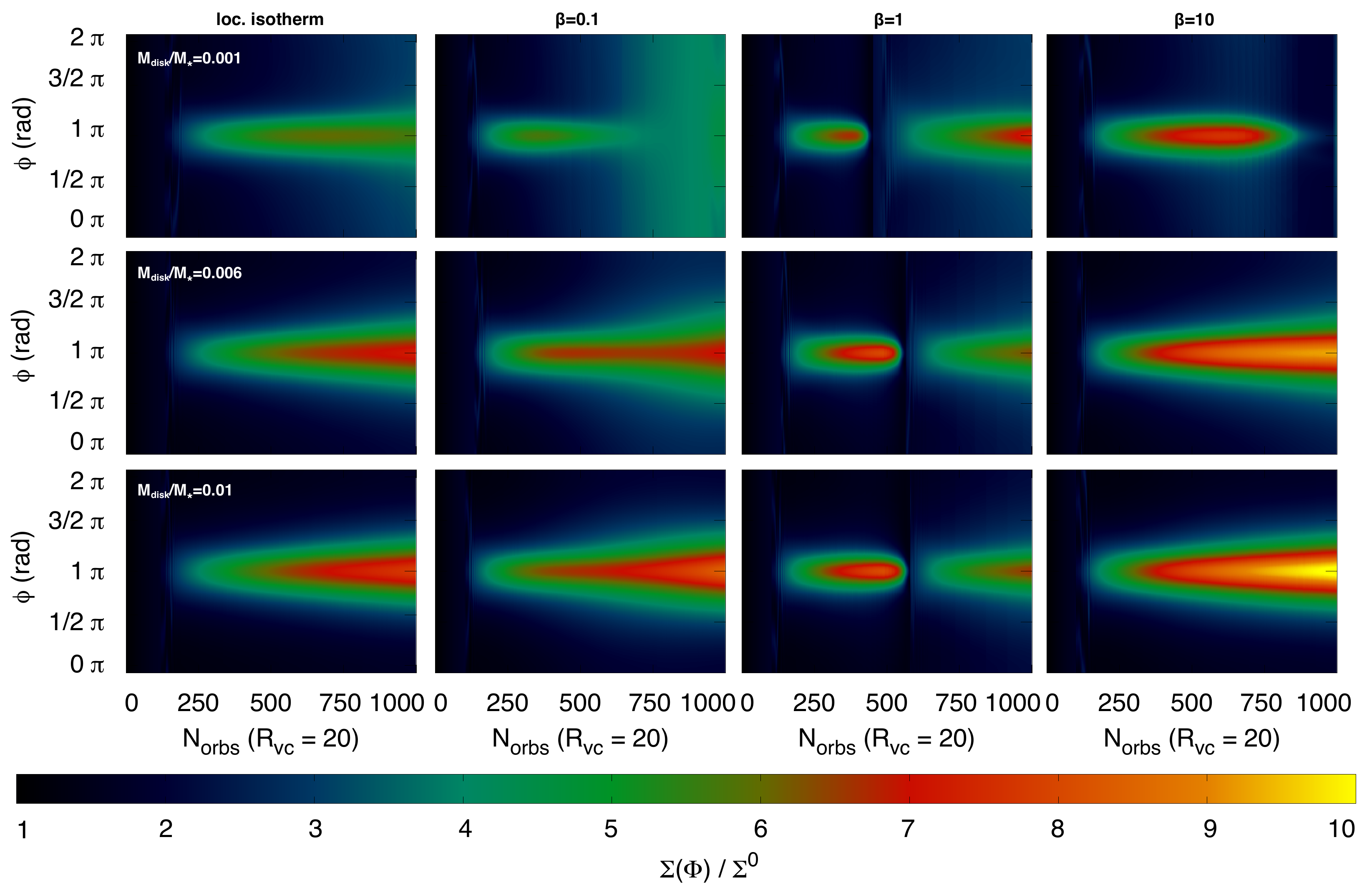}
    \includegraphics[width = \textwidth]{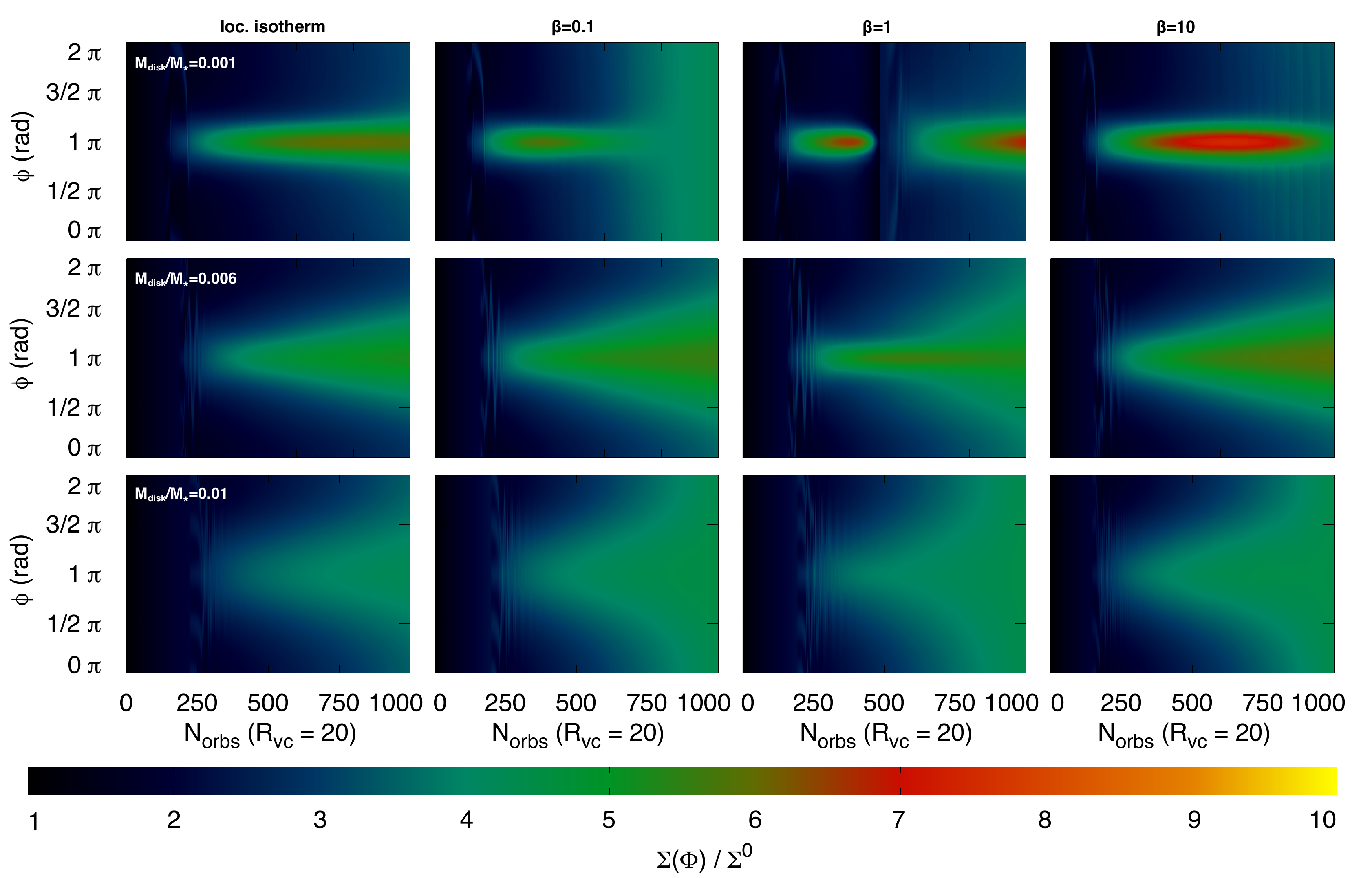}
    \caption{Vortex azimuthal density profile time evolution (in units of vortex orbit) without (upper panels) and with (lower panels) self-gravity, respectively. The first row refers to the low disk-mass models, the second row refers to the medium-mass models, while the third row is the case of high disc masses in Type~I simulations.
    From the left: first column are the locally isothermal cases, second column is the case when $\beta~=~0.1$, the third column is $\beta~=~1$ and the fourth column is $\beta~=~10$.}
    \label{fig:NOSGe-5_dens}
\end{figure*}

\begin{figure*}
    \centering
    \includegraphics[width = 0.97\textwidth]{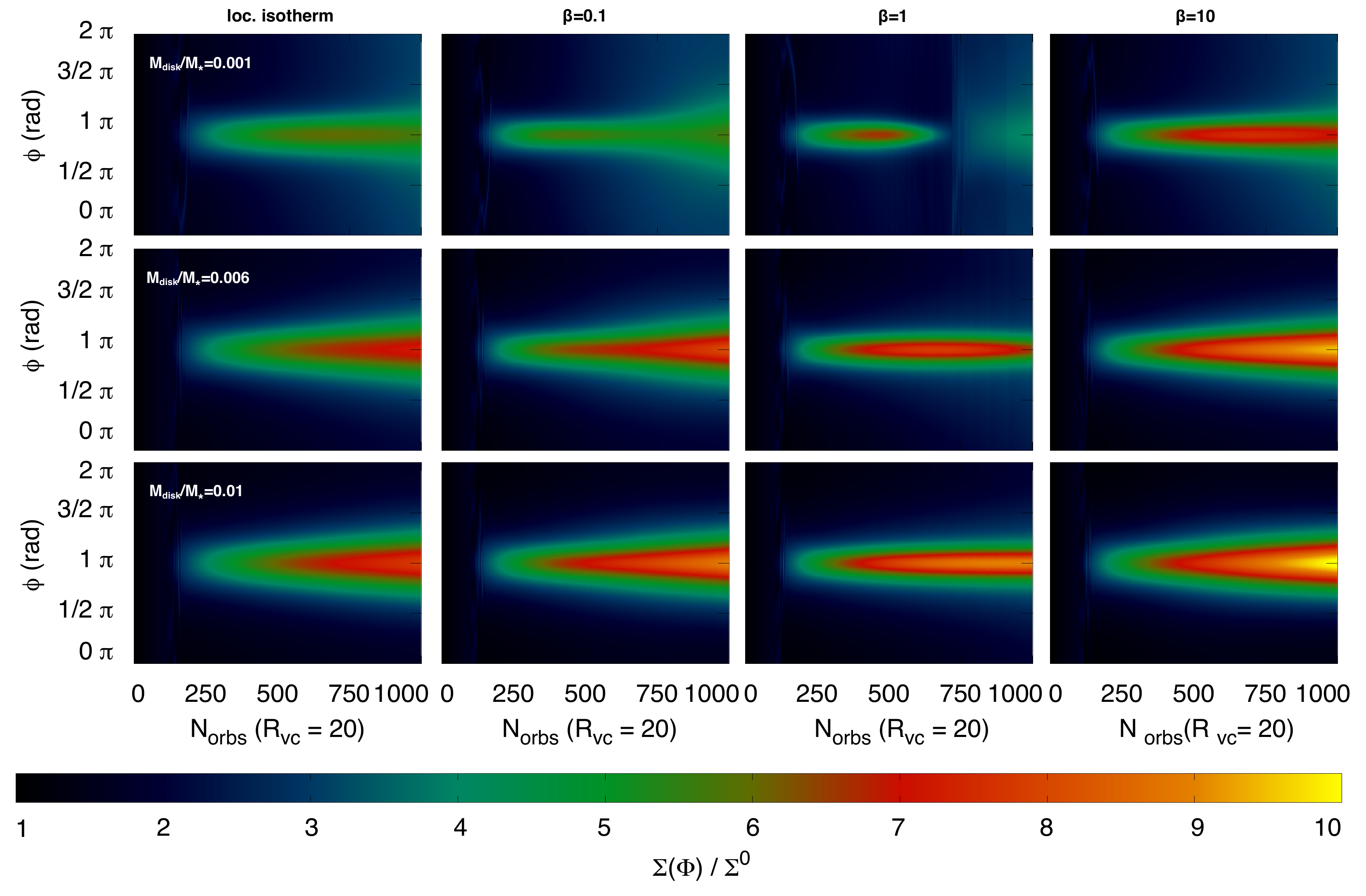}
    \includegraphics[width = 0.97\textwidth]{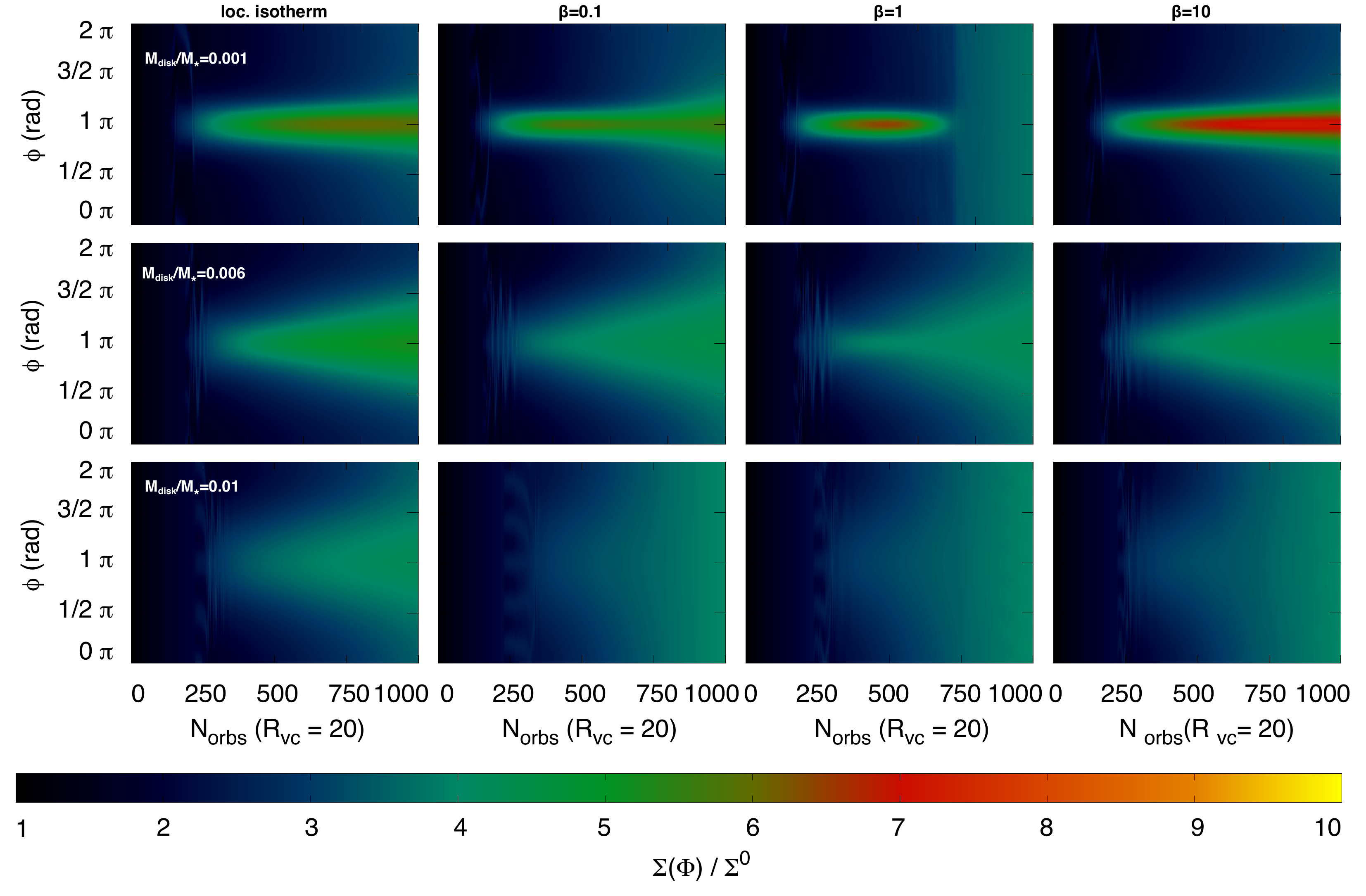}
    \caption{Vortex azimuthal density profile time evolution (in units of vortex orbit) without (upper panels) and with (lower panels) self-gravity, respectively. The first row refers to the low disc-mass models, the second row refers to the medium-mass models, while the third row is the case of high disc masses in Type~II simulations.
    From the left: first column are the locally isothermal cases, second column is the case when $\beta~=~0.1$, the third column is $\beta~=~1$ and the fourth column is $\beta~=~10$.}
    \label{fig:NOSGe-5_dens_T2}
\end{figure*}

A pressure maximum develops at the viscosity transition due to the accumulation of gas caused by the drop of radial transport there. We found that the RWI is excited at the pressure maxima, which results in the formation of small scale vortices with mode numbers $m~=~3-6$. Independent of the disc self-gravity, the RWI excitation is observed in all models. We found that the onset of RWI tends to occur at earlier times in non-isothermal models than in locally isothermal ones. As each small vortex forms at a slightly different distance from the central star, they tend to merge to form a single large-scale vortex. In the following, we present the evolution of the large-scale vortex in each investigated models.

We implicitly assume that the density distribution inside the vortex is elliptical as it is described by \cite{Kida1981} and \cite{Chavanis2000}. Theoretical vortex models have shown that a definite relation exists between the Rossby number, Ro (the rate of the rotation velocity of the vortex and the rotation velocity of the Keplerian disk), and the vortex aspect ratio (see details e.g. \citealp{Kida1981}, \citealp{GNG}, \citealp{SurvilleandBarge2015}). These authors have shown that the stronger the vortex (i.e., the larger the magnitude of the Rossby number), the smaller the aspect ratio is. 

To measure the vortex strength, first we calculate the vortex aspect ratio ($\chi_{\mathrm{dens}}$) in each model in the following way. On a polar grid, 2D elliptical contours are fitted to the surface density, normalised by the initial density, $\Sigma^0$. Then $\chi_{\mathrm{dens}}$, the ratio of the azimuthal and radial axes of the contour at 87\% of the maximum value of the normalised density is determined. Fig. \ref{fig:chi_contr_e-5} shows the evolution of $\chi_{\mathrm{dens}}$ in models without and with self-gravity on panels a-1) -- a-3) and b-1) -- b-3), respectively. By carefully analysing the Fig., one can conclude that vortices are full-fledged by 500th orbits in all models.

We also investigated the mean azimuthal density profile, $\delta \Sigma$, calculated at the vicinity of the radial distance of the vortex eye. First, we normalised the surface density with that of the initial. Then, the normalised density distribution is averaged radially, taking into account rings having $\pm~5$ cells radial distance centred on the maximum density. To obtain the evolution of the vortices, this procedure is done on each frame. Fig. \ref{fig:NOSGe-5_dens}  shows the time evolution (time is measured in units of the Keplerian orbit at the vortex distance) of $\delta \Sigma$-profiles in non-self-gravitating and self-gravitating cases, respectively.

 \subsection{Non-self-gravitating Type~I models}
 \label{sec:nosg}

Independent of disc thermodynamics, vortices tend to be stronger in higher mass models if disc self-gravity is neglected. In these models the minimum value of $\chi_{\mathrm{dens}}$ is always higher in the locally isothermal case than in the $\beta$-cooling models, see Panels a-1) -- a-3) in Fig. \ref{fig:chi_contr_e-5}. This means that vortices formed in the models where the thermodynamical effect is taken into account are less elliptical and therefore stronger.

In the low-mass models, the vortex lifetime strongly depends on the disc thermodynamics: the vortex life-time is shorter in the $\beta$-cooling models than in the locally isothermal model, see panel a-1) in Fig. \ref{fig:chi_contr_e-5}. A vortex decay and subsequent reappearance (steep growth in the value of $\chi_{\mathrm{dens}}$ followed by slow decay) can be seen for $\beta=0.3,\,1$, and 3. The reappearance of the large-scale vortex requires about 10 orbits, see upper panels on Fig.\,\ref{fig:NOSGe-5_dens}.

In high-mass models, the effect of thermodynamics on vortex evolution is less pronounced. In both the medium- and high-mass models, vortices are slightly stronger in the $\beta$-cooling models than in locally isothermal ones. As it can be seen on the left panels of Fig.\,\ref{fig:chi_contr_e-5}, vortex reappearance occurs at later epochs with increasing disc mass.

 \subsection{Non-self-gravitating Type~II models}
 \label{sec:nosgt2}
 
As described in Section \ref{sec:thermo}, we compared locally isothermal and non-isothermal simulations in Type~II models, where the initial aspect ratio differs, see Equation \ref{eq:h}. In this case, we assume that the initial temperature distribution is the same in locally isothermal and non-isothermal cases, see Equation~\ref{eq:csni}. Comparing Figs.~\ref{fig:NOSGe-5_dens} and \ref{fig:NOSGe-5_dens_T2}, one can see that change in the initial conditions affects the evolution of vortices. In Type~II models vortices live longer than in Type~I models independent of disc mass. Similar to Type~I simulations, we found that vortices weaken with increasing $\beta$. Moreover, in low disc mass simulations, the shortest vortex lifetime occurs for $\beta~=~1$ as in Type-I models. Note that, as we altered the aspect ratio of the non-isothermal models, locally isothermal simulations are the same as in the case of Type~I models.

\cite{PierensandLin2018} found that assuming sufficiently rapid cooling time ($\beta~\leq~0.1$), the results are consistent with the locally isothermal case (see Section~\ref{sec:intro}). Although, in this paper, we only focus on the effect of thermodynamics with higher $\beta$ values, to compare our simulations with their results, we also run simulations with $\beta~=~0.01$. Similarly, we found that in the case of such rapid cooling time, the disc tend to act as in the locally isothermal case, independent of disc mass. Fig.~\ref{fig:lowbeta} shows the time evolution of $\delta \Sigma$ of a low-mass non-self-gravitating Type~II model with $\beta\,=\,0.01$. Similar to the locally isothermal case, no vortex decay was observed. We also found that, in the case of rapid cooling time, $\delta \Sigma$ of the mature vortex is somewhat higher ($\delta \Sigma~\simeq~7$), than in the locally isothermal case ($\delta\Sigma~\simeq~6$, compare Fig.~\ref{fig:lowbeta} and the locally isothermal case of the low disc models in the upper panel of Fig.~\ref{fig:NOSGe-5_dens_T2}).

 \subsection{Self-gravitating Type~I models}
 \label{sec:sg}

In terms of vortex strength and evolution, the low-mass self-gravitating models are similar to non-self-gravitating models, see lower panels on Fig.\,\ref{fig:NOSGe-5_dens}. However, in higher mass self-gravitating models, significantly weaker vortices form, independent of disc thermodynamics. This can be explained by the vortex stretching effect of self-gravity described by \cite{RegalyandVorobyov2017a}. The stretching effect can be described by the gravitational torque (caused by the vortex). The leading (inner) part of the vortex loses angular momentum (as it suffers from negative gravitational torque). As a result, the inner part of the vortex moves faster to the central star, stretching the vortex inwards. In contrary, the outer part of the vortex is affected by positive gravitational torque. As a result, the outer part of the vortex gain angular momentum and moves outwards. This two opposite effects elongates and accelerates the decay of the vortex, see Fig.~6 of \cite{RegalyandVorobyov2017a} for more details.

Vortex reappearance occurs only in the smallest mass models. Contrary to the non-self-gravitating case, in medium- and high-mass discs with self-gravity we did not observe vortex reappearance, see lower panels on Fig. \ref{fig:NOSGe-5_dens}.

In all other cases, independently of disc thermodynamics the large scale vortex decays completely within 1000 orbits. Shortest vortex life-time is observed in $\beta=1$, while the longest vortex life-time is observed in the locally isothermal case, see panel b-3) of Fig.\,\ref{fig:chi_contr_e-5}. In these models, the stretching effect of self-gravity becomes dominant over thermodynamics. However, we observed a modest $\beta$-dependence of vortex life-time in medium-mass case, if $\beta=~0.3$ and $1$, see panels b-2) and b-3) of Fig. \ref{fig:chi_contr_e-5}.

\begin{figure}
    \centering
    \includegraphics[trim={5px 60px 10px 70px},clip,scale=.55]{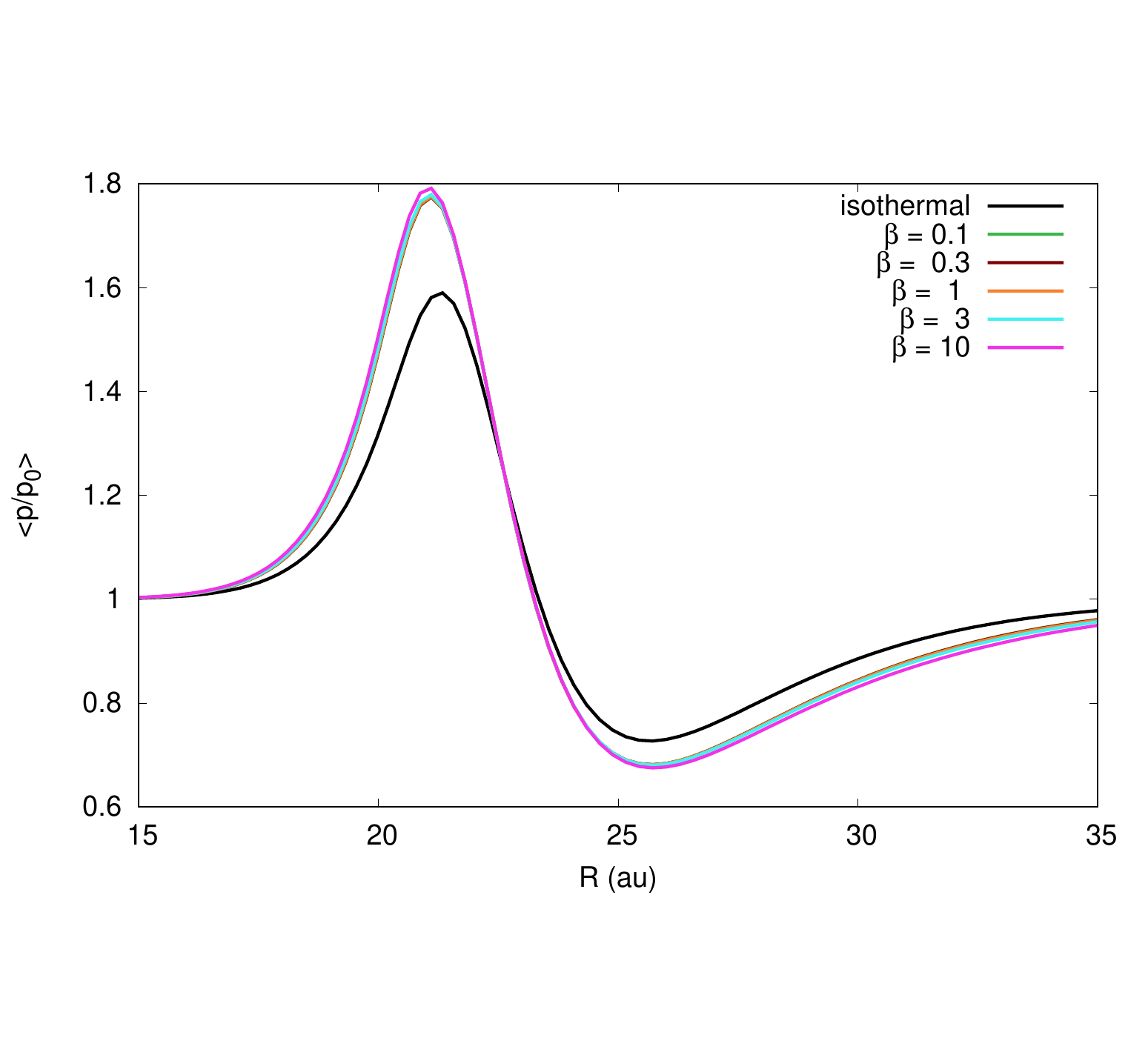}
    
    \caption{Evolution of the azimuthally averaged, radial pressure profile without self-gravity at $t\,=\,50$ orbits. The profile is normalised with the corresponding initial pressure profile. Y axis refers to the normalised pressure profile, while X axis refers to the distance from the star in astronomical units. The black solid line denotes to the isothermal case, the coloured lines refer to the non-isothermal cases.}
    \label{fig:1Dpress}
\end{figure}

 \subsection{Self-gravitating Type~II models}
  \label{sec:sgt2}

In Type~II simulations, the effect of self-gravity is less effective in low disc mass cases similarly to Type~I self-gravitating simulations. In $\beta~=~1$ case, the vortex dissipates in the less massive model, while both in $\beta~=~0.1$ and in locally isothermal cases, only weak vortices are formed. In $\beta~=~10$, the strength of the mature vortex is the highest. Similar to non-self-gravitating case, the shortest vortex lifetime is observed in low disc mass $\beta~=~1$ model. This phenomenon is caused by the same effect of the indirect potential observed in Type~I models.

\section{discussion}
\label{sec:discussion}

As described in Section \ref{sec:res}, we found that the onset of RWI excitation tends to occur earlier in non-isothermal cases, than in locally isothermal models. A possible explanation to this phenomenon is that the gas pressure maximum forms faster in the non-isothermal models. As a result, the pressure gradient can be steeper in the non-isothermal case.

The equation of state of the gas in the locally isothermal ($P_{\mathrm{I}}~=~\Sigma c_{\mathrm{s,I}}^2$) and $\beta$-cooling cases ($P_{\mathrm{NI}}~=~(1-\gamma)e$) are different. While $P_{\mathrm{I}}$ depends only on $\Sigma$ via $c_{\mathrm{s,I}}$ (see Section \ref{sec:hydro}), $P_{\mathrm{NI}}$ depends on the internal energy, which is governed by the energy conservation equation (see Equation~(\ref{eq:energy})). Thus, $\beta$ affects the onset of RWI via the energy equation in the non-isothermal case. Fig.~\ref{fig:1Dpress} shows a comparison of the radially averaged, normalised radial pressure profile at $t~=~50$ orbits, before the onset of RWI. We found that the pressure bump in the non-isothermal models are steeper than that in the locally isothermal models. The higher the $\beta$, the steeper the profile. Note, however, that the effect is weak.

We found that increasing $\beta$ leads to less elliptical, and therefore stronger vortices (see, e.g., Fig.\,\ref{fig:chi_contr_e-5}). Vortex strength can be measured by its aspect ratio ($\chi_{\mathrm{dens}}$). The strength of a vortex is maximum as $\chi_{\mathrm{dens}}$ reaches its minimum value. In all non-isothermal simulations, strongest vortex is observed for $\beta=10$. This is in agreement with \cite{PierensandLin2018}. However, they investigated models with $0.001 \geq \beta \geq 1$. Thus, our simulations revealed that for even higher $\beta>1$ this trend is valid. 

Additionally to the above described two effects of $\beta$-prescription, we found that increasing $\beta$ leads to shortened vortex life-time. The vortex evolution is governed by the Navier-Stokes equations which depends on local viscosity. Since we use $\alpha$-prescription, the kinematic viscosity of the gas depends on the local sound speed. As a result of the difference in the sound speed (see Equations~(\ref{eq:csi}) and~(\ref{eq:csni})) for locally isothermal and non-isothermal cases, vortices evolve on different time scales. While in a locally isothermal model the viscosity ($\nu_{\mathrm{I}}~=~\alpha c_{\mathrm{I}}^2/\Omega)$ is constant in time, in non-isothermal models, viscosity ($\nu_{\mathrm{NI}}~=~\alpha c_{\mathrm{NI}}^2/\Omega$) changes through $c_{\mathrm{NI}}~\propto~\sqrt{e/\Sigma}$ (see Equation~(\ref{eq:csni})). 

\begin{figure}
    \centering
    \includegraphics[width = \columnwidth]{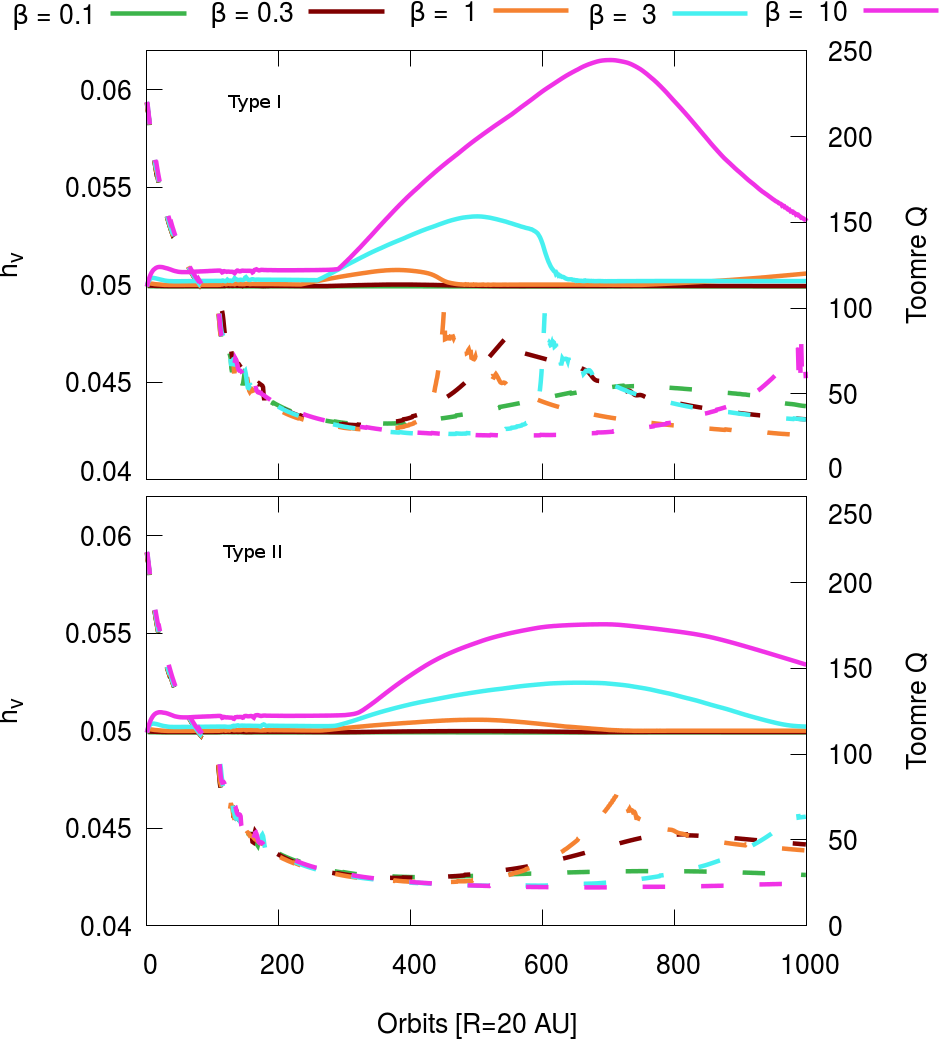}
    \caption{Evolution of the aspect ratio in the vortex eye ($h_{\mathrm{v}}$, solid lines) and the corresponding Toomre Q parameter (dashed lines) in non-self-gravitating Type I and Type II low mass non-isothermal models.}
    \label{fig:NOSGh}
\end{figure}

Moreover, as we assumed flat-disc approximation with $\alpha$-prescription of \cite{ShakuraandSunyaev1973}, accretion (driven by the kinematic viscosity of the gas), is connected to the the disk vertical scale height. Fig.~\ref{fig:NOSGh} shows the evolution of $h$ (solid lines) and the corresponding Toomre Q parameter (dashed lines) in the eye of the vortex in non-self-gravitating Type~I and Type~II models. It can be seen that increasing $\beta$ enhances the amplitude of variation in $h$. In $\beta\,=\,10$ model, at about 700 orbits $h$ is increased by about $10-20$\% with respect to the initial value (dependent of simulation type). As in Type~II simulations, we compared models with the same initial temperature profile, we changed the initial aspect ratio of the non-isothermal models (see Equation~\ref{eq:h}). As shown in the results of Type~II simulations, we found that the geometry of disc has also a substantial effect on vortex evolution and life-time.

\begin{figure*}
    \centering
    \includegraphics[width=\textwidth]{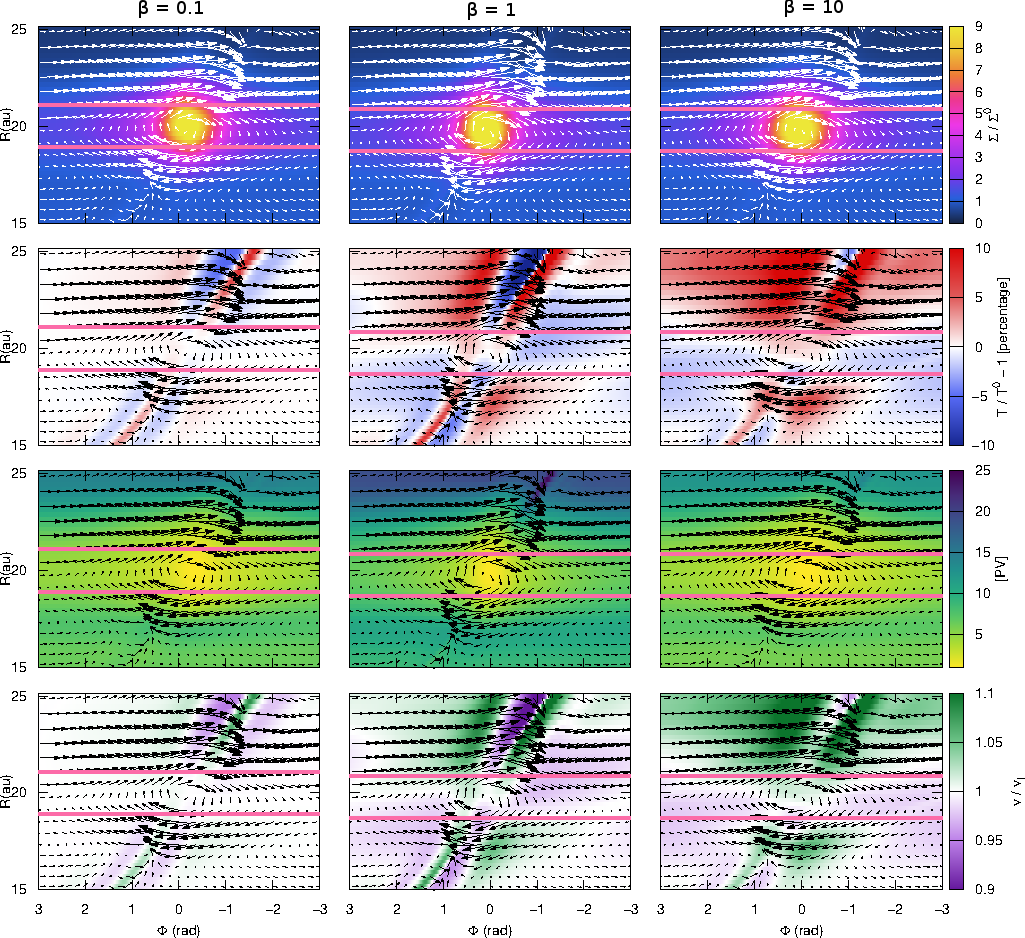}
    \caption{The normalised density ($\Sigma / \Sigma^0$), temperature ($T/T^0$), normalised potential vorticity ([PV]) and viscosity normalised to the isothermal one ($\nu / \nu_{\mathrm{I}}$) for high-mass ($M_{\mathrm{disk}}/M_{\star}~=~0.01$) non-self-gravitating disc model at $t~=~500$ orbits in Type~I simulations. From left to right the locally isothermal and $\beta~=~0.1,~1$ and $10$ models are shown. Arrows show the gas flow. Horizontal lines correspond to the zone where the radial averaging is calculated for density profiles.}
    \label{fig:NOSG_nu_M09}
\end{figure*}

\begin{figure*}
    \centering
    \includegraphics[width=\textwidth]{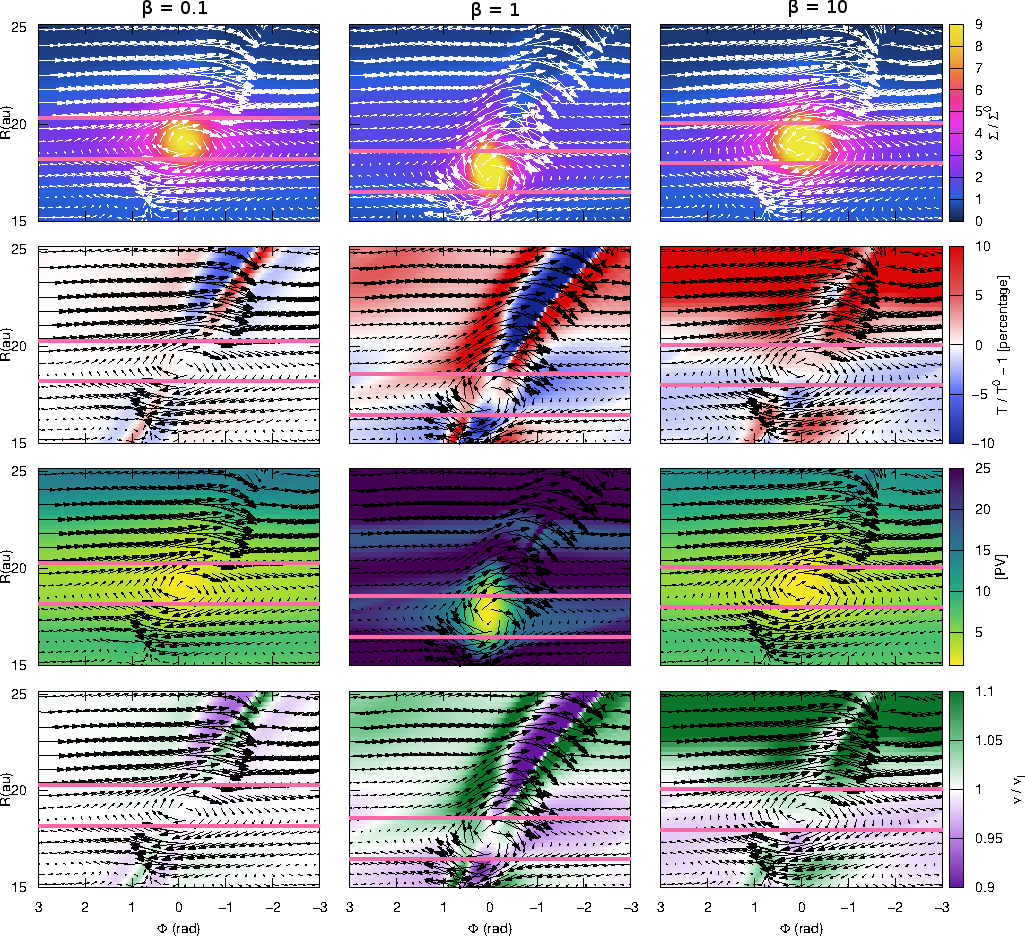}
    \caption{The normalised density ($\Sigma / \Sigma^0$), temperature ($T/T^0$), normalised potential vorticity ([PV]) and viscosity normalised to the isothermal one ($\nu / \nu_{\mathrm{I}}$) for high-mass ($M_{\mathrm{disk}}/M_{\star}~=~0.01$) non-self-gravitating disc model at $t~=~500$ orbits in Type~II simulations. From left to right the locally isothermal and $\beta~=~0.1,~1$ and $10$ models are shown. Arrows show the gas flow. Horizontal lines correspond to the zone where the radial averaging is calculated for density profiles.}
    \label{fig:NOSG_nu_M09t2}
\end{figure*}

Figs.~\ref{fig:NOSG_nu_M09} and \ref{fig:NOSG_nu_M09t2} present the normalised density with the initial one ($\Sigma/\Sigma^0$), temperature normalised by the initial one ($T/T^0$), the normalised potential vorticity (PV, $\zeta$, or vortensity) and viscosity normalised by the locally isothermal one ($\nu/\nu_{\mathrm{I}}$) in the non-self-gravitating $\beta$-cooling models at $t~=~500$ orbits. Left, middle and rigth panels represent $\beta~=~0.1,~1$ and $10$, in Type~I and Type~II simulations respectively. Arrows correspond the gas flow, while horizontal lines represent the zone within the radial averaging is calculated for $\delta \Sigma$.  Lower panels on Fig.~\ref{fig:NOSG_nu_M09} show that the viscosity in the eye of the vortex is about 5\% larger for $\beta=10$ compared to locally isothermal case in Type~I simulations. Note that in Type~II simulations, the ratio of the initial viscosity and the viscosity at 500 orbits are larger than in Type~I models (see Fig.~\ref{fig:NOSG_nu_M09t2}). This is because of the fact that as $\beta$ increases, the temperature ($T$) also increases inside the vortex (see middle panels of Fig.~\ref{fig:NOSG_nu_M09} and \ref{fig:NOSG_nu_M09t2}) due to slower cooling prescription. This leads to larger sound speed, which causes increased kinematic viscosity of the gas. 

Non-axisymmetric RWI occurs in the local minimum of vortensity (PV, \citealp[see e.g.][]{Lietal2000, Lietal2005, Kolleretal2003}), which is formed  on the edge of the dead zone in our simulations. Vortensity is measured as follows:
\begin{equation}
 \label{eq:pv}   
 \zeta = \frac{\vec{\omega}}{\Sigma}S^{-2/\gamma},
\end{equation}
where $\vec{\omega}~=~\nabla\times{\bf{\mathrm{v}}}$ is the vorticity (describing the curl of the velocity field), and $S~=~P/\Sigma^{\gamma}$ is the entropy. In Figs.~\ref{fig:NOSG_nu_M09} and \ref{fig:NOSG_nu_M09t2} vortensity is normalised by its minimum value ($\zeta_{\mathrm{min}}$). As one can see $\zeta_{\mathrm{min}}$ is located in the vortex eye in all cases.

As we have seen, the vortex lifetime is set by two competing effects of increasing $\beta$: strengthening of the vortex due to the steeper pressure gradient and weakening of the vortex due to the increased kinematic viscosity. These two effects can be clearly seen in the non-self-gravitating models (see Fig.~\ref{fig:chi_contr_e-5}). Independent of the disc mass, the shortest lived vortex is always found for $\beta~=~1$ models. $\beta<1$ leads to weaker, but longer lived vortices. Contrary, $\beta>1$ vortices are stronger, but suffer from faster vortex dissipation. 

We also found that the above strengthening and weakening effects can be seen in low-mass self-gravitating models, however, in middle and higher disc masses, the effect of thermodynamics becomes less dominant (see Fig.~\ref{fig:NOSGe-5_dens}). \citet{RegalyandVorobyov2017a} have shown that disc self-gravity tends to destroy vortices in locally isothermal cases, which also occurs in the non-isothermal discs.

\section{Conclusion}
\label{sec:concl}

In this work, we investigated the effect of thermodynamics on the long-therm evolution of RWI-excited vortices at the outer edge of the dead zone by means of 2D hydrodynamical simulations with three different disc masses ($M_{\mathrm{disk}}/M_{\star}~=~0.001, 0.006$, and $0.01$) in self-gravitating and non-self-gravitating models. The thermodynamical cooling and heating mechanisms are taken into account with the $\beta$-prescription. RWI excitation occurs inside the dead zone, where the viscosity is assumed to be low ($\alpha_{\mathrm{dz}}~=~10^{-5}$). In our simulations, the viscous heating is neglected, because vortices form inside the low-viscosity region where the viscous term is negligible. We investigated the effect of slow-, moderate- and rapid cooling/heating process, by assuming different $\beta$-parameters in the 0.1--10 limits. Rapid, moderate and slow cooling/heating mechanism occur on the timescales of $\tau_{\mathrm{c}}~=~0.1\Omega^{-1}$,$1\Omega^{-1}$and $10\Omega^{-1}$, respectively. We run Type I and Type II simulations. The former represents a comparison between non-isothermal and locally isothermal simulations with the same initial aspect ratio. This yields a difference between the initial sound speed (see Equation~(\ref{eq:css}). The latter provides a fairer comparison between the non-isothermal and locally isothermal simulations. For this, we set the initial sound speeds equal with reducing the initial aspect ratio of the non isothermal discs.
Our main findings are the following:

1) Thermodynamics affects the onset of RWI excitation through pressure gradients, which evolves on different timescales both in Type I and Type II simulations. In the non-isothermal case, the pressure gradient is steeper than in the locally isothermal case. In the locally isothermal case, pressure depends only on the surface density, while in the non-isothermal case, pressure depends on the energy, which is governed by the energy conservation equation, and hence by the $\beta$-parameter.

2) In the non-isothermal cases, less elliptical, therefore stronger vortices form than in the locally isothermal cases both in Type I and Type II simulations. Increasing $\beta$ leads to stronger vortices. 

3) $\beta$ has an effect on the vortex life-time via the kinematic viscosity of the gas. While in the locally isothermal model the viscosity is constant in time, in the $\beta$-cooling models viscosity evolves in time, affecting the evolution of the vortex via the Navier-Stokes equation. In the framework of $\alpha$-prescription, increasing $\beta$ leads to higher temperature, thus higher viscosity in the eye of the vortex, which leads to faster vortex dissipation.  This effect can be seen in Type I and Type II simulations as well.

4) In low disk-mass models ($M_\mathrm{disk}/M_*<0.006$), we observed similar effects of thermodynamics in self-gravitating and in non-self-gravitating models. This effect is independent of model type, it can be osberved both in Type I and Type II simulations. However, in higher mass discs ($M_\mathrm{disk}/M_*>0.006$), the vortex stretching effect of self-gravity becomes dominant over thermodynamics.

5) The effect of disc geometry ($h$) plays also a key role in vortex life-time and evolution. We found that decreasing the disc aspect ratio (by a factor of $\sqrt{\gamma}$ in Type II models) causes longer vortex life-time. 

Based on our results, we conclude that the lifetime of a vortex  is determined by two competing effects of thermodynamics: increasing $\beta$ strengthens vortices, while shorten their lifetime. The shortest vortex lifetime is found in models with $\beta~=~1$.

\cite{PierensandLin2018} showed that the details of thermodynamics are a crucial point in the long-term evolution of RWI-excited vortices. Because they concluded that at a pressure bump long-lived vortex can form in self-gravitating discs assuming black body cooling, this thermodynamical process is required to be included in our future work. 

We used two dimensional hydrodynamic simulations in the thin-disc approximation. \citet{LesurandPaploizou2009} showed that the vertical stratification in protoplanetary discs influence vortex evolution: for $\chi_{\mathrm{dens}}<4$ vortices are destroyed by the elliptical instability. For a better understanding of the effect of thermodynamics on the vortex evolution, a three dimensional hydrodynamic model with thermodynamics and self-gravity is needed. However, we note that our simulations may overestimate the effect of self-gravity compared to an equivalent 3D disk, as the gravitational softening is not included in our simulations.

We note that, we focused on simulations with $\beta\,\ge\,0.1$. Nonetheless, we also run simulations with $\beta\,=\,0.01$. We showed that in the case of sufficiently short cooling time-scales, the results of non-isothermal simulations are similar to the locally isothermal cases. This is in agreement with what was found by \cite{PierensandLin2018}.

We also note that we did not include the evolution of the dust in our simulations. However, \citet{Fuetal2014b} showed that vortices can effectively collect inward drifting dust particles. As the dust-to-gas ratio within the vortex starts to reach unity (or higher), the dust feedback destroys the vortex. Recently, \citet{Mirandaetal2017} showed that, although dust feedback affects the evolution of vortices, asymmetric dust accumulation can be observed in protoplanetary discs for thousands of orbits. As the dust feedback has a clear effect on the vortex evolution, more detailed simulations with thermodynamical processes are needed.

We found that increasing temperature increases viscosity. This phenomenon can be explained by the fact the we assumed a thin-disc  $\alpha$-prescription of \cite{ShakuraandSunyaev1973} ($\nu~\propto~c_{\mathrm{s,NI}}^2~\propto~T$). For investigating the effect of temperature on viscosity, a more realistic, fully magnetohydrodynamical (MHD) model is needed \citep[see e.g.][]{ZhuandBaruteau2016}.

Our results reveal the importance of thermodynamics in modelling protoplanetary discs as thermodynamics influences the evolution and lifetime of anticyclonic vortices. The vortex lifetime can be a crucial parameter in planet formation as the pressure maxima at the eye of vortices, being traps for solids, not only collects dusty material, but save them against stellar engulfment. We conclude that planet formation might be enhanced in cooler discs as vortices have longer lifetimes there.

\section*{Acknowledgements}
\addcontentsline{toc}{section}{Acknowledgements}

This project was supported by the Hungarian OTKA Grant No. 119993 and the Momentum grant No. LP2018-7/2019 of the Hungarian Academy of Sciences.
ZsR acknowledges support from the MTA CSFK Lend\"ulet Disc Research Group. EV acknowledges support from the Russian Science Foundation grant 17-12-01168. We gratefully acknowledge the support of NVIDIA Corporation with the donation of the Tesla 2075 and K40 GPUs. DTN acknowledges L. Kriskovics, L. Szabados and A. P\'al for their suggestions and helpful remarks. We also thank for the referee for the useful comments and remarks.








\appendix
\section{Solving the energy equation}
\label{sec:appenda}
\label{sec:substep3}

To Investigate non-isothermal discs we implemented a solver for polytopic equation of state in the 2D hydrodynamical \textsc{gfargo} code\footnote{http://fargo.in2p3.fr/-GFARGO-}, which solves the hydrodynamic equations (see Equations \ref{eq:cont}, \ref{eq:NS} and \ref{eq:energy}) with operator splitting method. This method breaks the partial differential equations (PDEs) into part which is a simplified approximation for the exact solution of the equations, but is more accurate than a single integration step based on old data \citep{StoneandNorman1992}.

The splitted parts in the solution are grouped into to steps, the source and the transport steps. In the transport step the equations represent the source and sink terms for each of the dependent variables (pressure, density, energy). This step is divided into three sub-steps. First, the code updates the velocities due to pressure gradients, gravitational forces and inertial forces (due to polar coordinates). Pressure gradient is calculated at the beginning of the step using the equation of state, the gravitational potential is computed from the Poisson equation, inertial forces are due to geometry (due to curvilinear coordinates) and acting on the momentum flux. 
The second sub-step uses the updated velocities to add the artificial viscous stress and dissipation and added to the momentum and energy equations. The third sub-step calculates the compressional/expansional heating/cooling term.

We implemented cooling and heating methods which take into account the viscous dissipation and and artificial $\beta$-cooling. According to \cite{StoneandNorman1992} the energy update can be given as

\begin{equation}
\frac{e^{n+1}-e^{n}}{\Delta t} = -p^{n+1/2}\nabla{\cdot \bm{v}} + Q_+ - Q_-,
\label{eq:cooling}    
\end{equation}
where $Q_+$ and $Q_-$ are the heating and cooling terms, respectively. As mentioned earlier in Section \ref{sec:thermo}, the effect of $Q_+$ is neglected in all simulation, as RWI excitation occurs in the low viscosity region. Here, the effect of viscous heating is low. By assuming that the time-centred pressure is $p^{n+1/2}~=~(p^{n}~+~p^{n+1})/2$, after trivial algebra one can get

\begin{equation}
\label{eq:enup}
e^{n+1} = \frac{\left[ 1-(\Delta t/2)(\gamma - 1) (\nabla{\cdot \bm{v}}) \right]e^{n} - \Delta t Q_-   }{1+(\Delta t/2)(\gamma - 1) (\nabla{\cdot \bm{v}})}.
\end{equation}

Using the $\beta$-cooling prescription of \citet{LesandLin2015}, the cooling term is defined as
\begin{equation}
Q_- = \frac{1}{\tau_{c}} \left ( e^{n} - e^0\frac{\Sigma^{n}}{\Sigma^0} \right ),
\label{eq:q-les}    
\end{equation}
where $e^0$ and $\Sigma^0$ are the surface mass density and thermal energy density at $t = 0$, and  $\tau_c~=~\beta \Omega^{-1}$ is the cooling time.  With this Equation~(\ref{eq:enup}) can be written as

\begin{equation}
\begin{split}
e^{n+1} = \left [\frac{1 - (\Delta t/2)(\gamma - 1) (\nabla{\cdot v})}{1 +(\Delta t/2)(\gamma - 1) (\nabla{\cdot v})} \right ]e^n  - \\ -\frac{\Delta t}{\tau_c} \left [  \frac{\left ( e^n - e^0 \Sigma^n / \Sigma^0\right )}{1 +(\Delta t/2)(\gamma - 1) (\nabla{\cdot v})} \right ]
\end{split}
\label{eq:coolenergy}    
\end{equation}


\bsp	
\label{lastpage}
\end{document}